\documentclass[12pt]{article}%
\usepackage{amssymb}
\usepackage{amsmath}
\usepackage{amsfonts}
\usepackage{sw20ssur}
\usepackage{graphicx}
\usepackage{accents}%
\providecommand{\U}[1]{\protect\rule{.1in}{.1in}}
\newtheorem{theorem}{Theorem}

\newtheorem{proposition}[theorem]{Proposition}

\ifx\pdfoutput\relax\let\pdfoutput=\undefined\fi
\newcount\msipdfoutput
\ifx\pdfoutput\undefined\else
\ifcase\pdfoutput\else
\ifx\paperwidth\undefined\else
\ifdim\paperheight=0pt\relax\else\pdfpageheight\paperheight\fi
\ifdim\paperwidth=0pt\relax\else\pdfpagewidth\paperwidth\fi
\fi\fi\fi
\begin{document}
\begin{center}
\textbf{The Big Bang\thanks{\ For reasons made clear in the paper, the phrase "Big
Bang" is used (as in its original meaning) to mark the very beginning of the
Universe. The initial decelerating, matter-creating period (here) retains its
original name "reheat" instead of Big Bang. } : Origins and initial conditions
from Self-Regulating Cosmology (SRC) model}
\end{center}
\hfill\\
Manasse R. Mbonye$^{1}$\\
\hfill\\ 
$^{1}$ Department of Physics, College of Science and Technology, University of Rwanda, Rwanda
 \\ \\\\
Correspondence:mmbonye@gmail.com\;\;\;\;\;\;\;\;\;\;\;\;\;\;\;\;\;\;\;\;\;\;\;\;\;\;\;\;\;\;\;\;\;\;\;\;\;\;\;\;\;\;\;\;\;\;\;\;\;\;\;\;\;\;\;\;\;\;\;\;\;\;\;\;\;\;\;\;\;\;\;\;\;\;\;
\begin{center}
\textbf{Abstract}
\end{center}
Generating appropriate initial conditions for the Universe is key to
discussing cosmic evolution constructively. In standard cosmology the
traditional approach assumes an early Universe that emerges from an infinite
density, spacetime singularity. It then undergoes inflationary expansion,
followed by a matter-creating "reheat" period. This approach produces results
generally in good agreement with observations. However, to date it is not
known how (or even whether) a true past-directed spacetime-singularity can
generate a regular spacetime that becomes the observed Universe. It has
previously been suggested that appropriate initial conditions should emerge
naturally from existing physics rather than be set a priori. In this work we
seek to generate initial conditions predicated on the Self-Regulating
Cosmology (SRC) model recently presented [1]. In SRC, cosmic dynamics leads to
a universe that also self-regenerates from one evolutionary \textit{cosmic
phase} or (hereafter) \textit{kalpa}\textbf{\footnote{Wiki: \textit{Kalpa
(Sanskrit lit): a long period of time related to the lifetime of the universe
(creation).}} }to another. Within each such kalpa the dynamics evolves between
two disparate scales of horizons. The end of a kalpa and the beginning of the
next interface through a phase transition. The interface of the two phases
sets up the phase initial conditions of the emerging phase. These include: a
natural time-reset, and energy, volume (space) and entropy conditions. The
issue of whether (or not) the early Universe undergoes inflation becomes
naturally self-manifest in this treatment. In the context of the SRC model, we
discuss the Initial Entropy Problem (IEP). Lastly, we highlight on current
observations of over-mature (galaxy and blackhole) structures at high
redshifts, by JWST.
\\
\hfill\\
\section{Introduction}\label{introduction}
Modern cosmology traces the origins of the Universe back to a singular event
of infinite density and pressure, named the Big Bang by Hoyle and considered
to be a space-like singularity [2]. Spacetime singularities, including this
initial one in cosmology, arise as solutions of General Relativity to problems
involving strong gravitational fields. In the sixties, research in this area
led to statements now known as the Hawking-Penrose singularity theorems which
described features of singularities in black holes (future-directed) [2] and
in cosmology (past-directed) [3]. While theories of gravity predict spacetime
singularities, including the initial singularity [4], it is also the case that
no observational evidence for physical singularities exits, yet. And in the
recent past there have been inquiries on whether, or not, gravitational
collapse can avoid them \textbf{[5]}. In line with such inquiries non-singular
spacetimes have also been proposed in black holes (see for example [6] [7] [8]
[9]) and in cosmology (see for example [10] based on regular matter turning
into repulsive fluids with appropriate equations of state at high densities.
Other efforts to smooth out the initial singularity using quantum approaches
have included the famous Hartle-Hawking no-boundary approach [11] and the
tunneling approach by Vilenkin.[12]. There have also been discussions [13]
questioning whether such quantum approaches do indeed provide a satisfactory
smooth beginning for cosmic spacetime. A comprehensive review of quantum
applied methods to smooth out singularities can be found in [14].

In the 1980s the inflationary paradigm was introduced in cosmology to address
contemporary issues that included the horizon problem, the flatness problem
and the monopole problem [15]. It was even hoped then that a similar approach
could also possibly smooth out the initial singularity [16]. While the
inflationary paradigm was able to address several of these problems it does
not eliminate the singularity problem; though it merely pushes it back to a
time before inflation. The current most accepted model of modern cosmology,
$\Lambda CDM$, which has been quite successful interpreting observations [17],
incorporates this early era including of inflation. As a result, among some of
its challenges [18], the model still faces the unexplained physics of a
singularity-sourced cosmic origin, and how cosmic dynamics is generated and
proceeds thereafter with no known causal mechanism to generate inflation.
Without a doubt, understanding initial conditions for the Universe is key to
discussing cosmic evolution accurately and predictively. It could also
potentially aid in addressing better some of the other issues such as, the
initial entropy problem (IEP) [19], and the origin of primordial fluctuations
[20], to name a few. The challenge here is how to rid spacetime of the initial
singularity and thereby infer what led to inflation if indeed there was one,
or else seek alternative consistent initial cosmic evolution. Several
cosmological models have lately been constructed, seeking to make improvements
in the existing standard model of cosmology. They include, Conformal Cyclic
Cosmology (CCC) [21], the Ekpyrotic model [22] [23], and the Periodic\ model
[24], to name a few.
In this paper we present different Big Bang initial conditions in the context
of a recently presented model of Self-Regulating Cosmology (SRC) [1]. In SRC,
the self-regulating cosmic dynamics leads the Universe into evolutionary
phases of self-regeneration, here called kalpas. As we shall find, the end of
a given kalpa and the beginning of the next are linked by an interface in form
of a phase transition, between two states. We describe evolutionary features
from the preceding cosmic phase leading to this phase transition using a
simple language that makes use of known emergent processes as analogues, for
intuition. The transition process, referred to as the Big Bang for the new
kalpa, sets initial conditions for this, latter phase. The issue of whether or
not there is an inflationary era associated with the Big Bang\textit{ }finds a
natural explanation in this scenario. We close with a discussion on the issue
of mature structures observed by JWST at high redshifts [25]. The rest of this
paper is set up as follows. In Section 2 presents a brief overview of the SRC
model. In Section 3 we make a case for the Big Bang based on SRC, and uses it
to present generate cosmic initial conditions in section 4. In section 5 we
comment on expected initial dynamical evolution while in Section 6 we point to
JWST observed mature structures at high redshifts as a potential observational
test of the model. Section 7 concludes the discussion.

\section{ SRC Model: Overview}

In this section we provide an overview of the SRC model. In General
\ Relativity, the Einstein Field equations imply components of the energy
momentum tensor (such as matter-energy density $\rho$ and pressure $p$) induce
dynamics on space. In particular, in Friedman cosmology (where the
Cosmological Principle is assumed) cosmic dynamics is described by an elegant
equation, the Friedman second equation $\frac{\ddot{a}}{a}+\frac{4\pi G}%
{3}\left(  3p+\rho\right)  =0$, where $a\left(  t\right)  $ size-scales the
non-static universe. The model of Self-Regulating Cosmology (SRC) arises out
of consideration that cosmic dynamics is (actually) sourced from two separate
(albeit related) drivers. The first such driver is space-time, itself, (for
this purpose, hereafter referred to as basic-space). In the model, basic-space
is posited to emerge from some underlying energy-based structure [1]. As we
show below, under ideal or unperturbed conditions, basic-space structure is
taken here to satisfy a net-zero gravitational charge (or net-zero-mass)
condition: $3p_{s}+\rho_{s}=0,\ (\rho_{s}\neq0,\ p_{s}\neq0)$. The result is
that $\ddot{a}_{s}=0$. Accordingly, a universe free of net matter-fields is in
dynamic equilibrium and will still expand, albeit with a coasting evolution
$a_{s}(t)\propto t$. The model further, posits that such dynamic equilibrium
state can be disturbed by energy-based perturbations (e.g. quantum field
fluctuations), with the result that basic-space curves and/or stretches in
response, consistent with the demands of General Relativity. This leads to the
second driver of cosmic dynamics, namely free matter-energy
fields\textit{\footnote{The phrase "free matter-energy fields" here includes
all fields (other than the fabric of basic-space depicted in (Eqs. 2.1 to
2.4)).}}. In the model the Universe's motivation for matter-energy creation is
an apparent need to offset shifts from cosmic dynamic equilibrium. It is in
this sense that cosmic dynamics becomes inherently self-regulating. These
characteristics and their motivation \ form the underpinnings of the SRC model
as summarized in a 3-proposition statement (below), referred to as the Dynamic
Equilibrium Protection Proposal (DEPP):

\begin{proposition}
An ideal universe, constituted purely by basic-space, remains in a state of
constant (coasting) expansion unless perturbed by free matter-energy fields.
\end{proposition}

\begin{proposition}
The Universe creates dynamic stability against perturbations that tend to
shift it from its coasting, dynamic equilibrium state. (Such can include
density perturbations that can grow from quantum fluctuations)\textit{ }
\end{proposition}

\begin{proposition}
\textit{When perturbed, the Universe will suppress the perturbations through
creation of free matter-energy fields with a net gravitational charge opposite
that of the perturbations.}
\end{proposition}

In this work we show that in addition to its self-regulating characteristic,
such a universe is also self-regenerating, going through periodic phases (here
called kalpas), with each phase commencing with a phase-transition (big bang).
Such features imply a model which introduces additional features to the
standard model. A full derivation of the SRC framework can be found in [1]. In
the current work, we discuss how the SRC model sets conditions for (and
predicts) the Big Bang naturally. This, in turn, leads to the observed cosmic
evolution that follows the big bang thereafter.

The above DEPP propositions imply that the classical (and hence emergent)
features of \textit{elasticity-like }expansion of basic-space, as a response
to free matter-energy fields, betray existence of an underlying fundamental
reality of its structure. In SRC [1] this emergent net-zero-mass basic-space
is modeled as a perfect fluid whose structure has non-vanishing density
$\sigma_{s},$ and pressure $\hat{\pi}_{s}$, representable as a diagonal
stress-energy tensor, $T_{\nu}^{\mu}=diag(\sigma_{s},-\hat{\pi}_{s},-\hat{\pi
}_{s},-\hat{\pi}_{s})$ [1], [26]. Noting that in absence of free matter-energy
fields the system resides in dynamic equilibrium so that $\ddot{a}_{s}=0$, and
setting spatial curvature $k=0$ (also consistent with persisting observations
[27]) the Einstein equations for the density and pressure evolutions give, respectively,%

\begin{equation}
\frac{\dot{a}_{s}^{2}}{a_{s}^{2}}=\frac{8\pi G}{3}\sigma_{s}, \tag{2.1}%
\end{equation}
and
\begin{equation}
\frac{\dot{a}_{s}^{2}}{a_{s}^{2}}=-8\pi G\hat{\pi}_{s}, \tag{2.2}%
\end{equation}
Eqs. 2.1 and 2.2 (or even the 2nd Friedman equation above) imply basic-space
that emerges from the underlying energy-structure, satisfies an associated
equation of state of the form\textit{\footnote{The net-zero-mass equation of
state has also been applied in literature, even in situations involving
free-matter fields (see e.g. [27])}}%

\begin{equation}
\hat{\pi}_{s}=-\frac{1}{3}\sigma_{s}\text{.} \tag{2.3}%
\end{equation}
These features induce on spacetime [1] a special Lamaitre-Robertson-Walker
(L-R-W) metric
\begin{equation}
ds^{2}=c^{2}dt^{2}-a_{s}\left(  t\right)  ^{2}\left(  dr^{2}+r^{2}d\Omega
^{2}\right)  , \tag{2.4}%
\end{equation}
where $c$ is the speed of light. The scale factor is linear in time
$a_{s}\left(  t\right)  \propto t$. Thus the background basic space will
induce a coasting or linearly expanding universe (see also [27]), consistent
with Proposition 1 in DEPP.

Further, the model posits that, to lowest scale, the dynamics of the Universe
that takes consideration of\ superimposed effects of free matter fields will
then take the form $a\left(  t\right)  \sim a_{s}\left(  t\right)  \ +\delta
a\left(  t\right)  ,$where $\delta a\left(  t\right)  $ is determined from the
contribution to $a\left(  t\right)  $ due to the free matter fields [1]. In
highlighting the determine the contribution $\delta a\left(  t\right)  $ due
to free matter fields (see [1] for details) one starts by considering a
fluid-flux of number density $n$, four-velocity $u_{\mu}$, satisfying
$n_{;\mu}u^{\mu}+n\Theta=n\Gamma$, with a generating source of rate $\Gamma$
and expansion $\Theta$. For an adiabatic case one finds $\sigma_{s}%
+3H(\sigma_{s}+\pi_{s}+P_{c})=0$, where $P_{c}=-\frac{1}{3H}\left(  \sigma
_{s}+\pi_{s}\right)  \Gamma$ is identified as the creation pressure [28].
Using thermodynamic arguments [28][29] and Eqs. 2.1 and 2.2 along with DEPP
\textbf{[1]} one finds (see [1]) that $\delta a\left(  t\right)  $ takes on a
harmonic form
\begin{equation}
\delta a\left(  t\right)  =A_{m}\sin\left(  \omega t+\psi\right)  , \tag{2.5}%
\end{equation}
with $A_{m}$ as the oscillation amplitude and $\psi$ as an arbitrary phase angle.

Considering appropriate conditions [1] one finds the resulting scale factor
$a\left(  t\right)  \sim a_{s}\left(  t\right)  \ +\delta a\left(  t\right)  $
for cosmic dynamics to be given by%

\begin{equation}
a(t)=\upsilon^{\ast}\left[  t+\tilde{\tau}\sin\left(  \frac{t}{\tilde{\tau}%
}+\psi\right)  \right]  , \tag{2.6}%
\end{equation}
where $\upsilon^{\ast}$ is a normalizing factor with dimensions of inverse
time, $\tilde{\tau}=\frac{\tau}{2\pi}=\frac{1}{\omega}$ is a (reduced) period
(a new cosmological parameter to be determined) and $\psi=\frac{2\pi t_{f}%
}{\not \tau }$, the phase angle with $t_{f}$ (fundamental interval) later
fixed by specific boundary conditions to be of order of Planck time.$t_{f}\sim
t_{p}$. Further, one finds this $a\left(  t\right)  $ to satisfy a general
evolutionary equation
\begin{equation}
\tilde{\tau}^{2}\ddot{a}+a(t)=f(t) \tag{2.7}%
\end{equation}
where, here $f(t)$ is a particular solution to Eq. 2.6 that satisfies
$f(t)=\upsilon^{\ast}t$. It is the contribution to cosmic dynamics from
basic-space. From Eq. 2.6, the corresponding Hubble parameter takes the form
\begin{equation}
H(t\mathbf{)=}\frac{1+\cos\left(  \frac{t}{\tilde{\tau}}+\psi\right)
}{t+\tilde{\tau}\sin\left(  \frac{t}{\tilde{\tau}}+\psi\right)  }\mathbf{.}
\tag{2.8}%
\end{equation}
Eqs. 2.6 to 2.8 describe cosmic evolution in the SRC model.

The SRC model has been shown to posses several desirable advantages. These
include the its ability to reproduce the correct values of the observed
parameters, based on a minimum of assumptions. In particular, SRC can account
for the main dynamic evolutionary features of the Universe at any time in the
past to the present and predict details of future evolution. For this
dynamical evolutionary account, SRC requires from observations constraints on
essentially 2 free parameters. These parameters are $t_{ac}$, the look-back
time to when the observed current cosmic acceleration commenced, and $d_{0}$,
often called the current dimensionless age [30]. gives the product of the
current Hubble parameter and the age of the Universe, \textit{ }$d_{0}%
=H_{0}t_{0}\ $\textit{\footnote{Note here that theoretical model needs only
$d_{0}$\ as a single parameter to fix current $H_{0}$ and $t_{0}\ $predict and
$H(t)$ any time during entire cosmic dynamics.}. }With these, the model
independently determines, separately the current age $t_{0}$, the current
Hubble parameter $H_{0}$, and this parameter's general time evolution $H(t)$
in general. Additionally it gives the relative magnitudes of matter and dark
energy density parameters, $\Omega_{m}\ $and $\Omega_{de}$. And finally, the
process the model fixes $\tau$, the period of one evolutionary phase or kalpa,
which is a new cosmological parameter unique to SRC. The third free parameter
$\psi$, only needed to fix the initial conditions (and is shown to generally
make very little contribution at any other time as will be evident in the next
Section). Therefore the model requires constraint of, at most, only 3
parameters to describe cosmic dynamics. Observations [31] indicate $t_{ac}%
\sim4Gyr$. Therefore taking $t_{ac}=4.07Gyr$ and based on [32] setting the
current dimensionless age estimate of $d_{0}=0.96\ $[32], we have found [1]
for the Universe, the corresponding age to be $t_{0}=13.77Gyr$, the current
Hubble parameter to be $H_{0}=68.22km/s/mpc$ and the corresponding matter and
dark energy parameters to be $\Omega_{m}=30.1\%$ and $\Omega_{de}=69.9\%$,
respectively. These results are consistent with the most recent observations,
[33], [34]. The model also reproduces the general cosmic evolutionary profile,
consistent with the early matter dominating phase\ and which evolves into the
latter dark energy dominating phase. Overall these results indicate that,
amazingly, the SRC reproduces the observed cosmology and indicate the model
should be taken seriously. SRC, further, makes predictions about the future
evolution of the Universe, including a graceful exit from the current
acceleration and its future fate. The Standard model does not explain how the
current acceleration ends, or the fate of cosmic evolution thereafter other
than the occurrence of a big rip. In the SRC model the Universe expands
monotonically, while\ at the same time oscillating about a coasting path.
Figure 1 sketches the time evolution of the scale factor $a\left(  t\right)  $
depicted in Eq. 2.6.%

\begin{figure}[ptb]%
\centering
\includegraphics[
height=2.0358in,
width=3.0519in
]%
{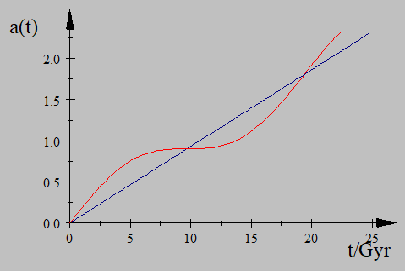}%
\caption{Figure 1 \textit{shows the general features of the time evolution of
the scale factor }$a(t)$\textit{, as depicted by the SRC model. The net
expansion (red) which is a combination of two independent influences
(basic-space and free matter-energy) is shown as an oscillatory curve about
the linear (equilibrium) expansion rate of basic-space. Here we have used
}$\tau_{\frac{1}{2}}=9.69Gyr$\textit{\ as the working cosmic half period.
Note: }$a_{0}\ =1$ at $t=13.76Gyr$\textit{.}}%
\end{figure}

One of the distinguishing features of the Universe as predicted by SRC (see
Figure 1) is self regeneration or self re-initiation at the end of each phase
$\tau$. In our discussion we shall consider this period $\tau$ to commence at
a Big Bang (such as that predicted by SRC to have occurred $13.77Gyrs$ ago).
The periodic evolution then extends through the decelerated structure
formation phase, into the current accelerated phase and to the latter's
predicted end, at which point another Big Bang ensues. For the constraints
$t_{ac}=4.07Gyr$ and $d_{0}=0.96$ the cosmic periodic phase (a new parameter)
computes to $\tau=19.36Gyrs\ $[1].

In the rest of this work we show how the end of each such a cosmic period
$\tau$ culminates into a phase transition that becomes the big bang, which
initiates the next cosmic phase or kalpa. We model and discuss the transition
from the past cosmic phase to a contemporary one, as an example of a big bang
in SRC. We demonstrate the non-singular origins of this phase transition (and
hence, generally, the non-singular origins of the universe). Lastly, in the
context of SRC, we highlight on other cosmic issues such as the Initial
Entropy Problem (IEP), the Cosmological Constant Problem, and the issue of
mature structures currently observed in early Universe by the James Web
Telescope (JWST).

\section{Making a case for the Big Bang}

In SRC cosmic dynamical evolution is a monotonic expansion (see Eq. 2.6,
Figure 1) represented by a 2-term scale factor $a\left(  t\right)
=a_{s}\left(  t\right)  +\delta a\left(  t\right)  $, with a background
contribution $a_{s}\left(  t\right)  $ from basic space and a harmonic
contribution $\delta a\left(  t\right)  =a_{m}(t)$\ arising from free matter
fields. In the model, the latter constitutes a perturbation on the background
former. The resulting motion traces out a periodic trajectory about a constant
gradient expansion path, $a_{s}\left(  t\right)  \propto t$. This motion (see
DEPP in section 2) implies a universe with self-regulating
characteristics,\ one always in search of a dynamic equilibrium state. Thus,
in an idealized situation in which matter fields are absent the Universe would
maintain a lasting state of dynamic equilibrium, expanding along that same
coasting path. Beyond its self-regulation character, such a universe is also
self-regenerating, with a specific self-regeneration interval $\tau$ [1]. In
this section we apply these features of SRC to the observed Universe to seek
for, set up and study appropriate, non-singular, initial conditions that can
be consistent with observations. One expects such initial conditions will turn
out to be those that describe features and events initiating self-regeneration
at a kalpa-to-kapla interface. The current aim is therefore to link features
and events that mark the end of one cosmic kalpa with those that mark the
commencement of the next one.

We start here by generally identifying the current cosmic phase or kalpa (see
Figure 1), whose initial conditions we seek, as some $k^{th}$ kalpa. Here, the
counting index $k$ can take on any positive integral values $k=\left\{
1,2,....\right\}  $ and its current value is arbitrary (and therefore not any
more special for our considerations). We denote by $t_{k}$ the time evolution
during the $k^{th}$ kalpa such that $0\leq t_{k}\leq\tau$. In general, each
such cosmic phase is, in turn, bounded by 2 others: one before and the other
after. Thus, for example, bounding the $k^{th}$ kalpa (not shown in Figure 1),
are the immediate past $\left(  k-1\right)  ^{th}$ kalpa, whose time evolution
is denoted by $0\leq t_{k-1}\leq\tau$ and the immediate future $\left(
k+1\right)  ^{th}$ kalpa, with a time evolution denoted by $0\leq t_{k+1}%
\leq\tau$. Specifically, the consideration here will (with no loss of
generality) focus on interfacing the end of the $\left(  k-1\right)  ^{th}$
kalpa and commencement of the $k^{th}$ kalpa. This interfacing will give rise
to the very initial conditions we seek for our current cosmic phase, here
referred to as the $k^{th}$ kalpa. It is in terms of the time-evolution,
within such time interval $\tau$ during a given phase, that both the local
dynamic and thermodynamic evolutions of the Universe can be described.

We remark that the time-identifications here, such as $t_{k}=0$, are
essentially just coordinate times which simply conveniently align with the
harmonic evolutionary functions in Eqs 2.5-2.8. The crucial exercise of
setting up initial conditions on \textit{physical time}\footnote{We will coin
the phrase "\textit{physical time}" here to refer to either "proper time" or
"comoving time" as opposed to coordinate time and conformal time.} flow, as
one of features above that mark the commencement of the next kalpa, is done
later in this section. As we find, this exercise will require identifying a
physical feature or event that justifiably relates to initiating physical time
flow. The exercise will be analogous to how temperature, as a measure of
average kinetic energy of particles, is initiated in the Kelvin (absolute)
scale by the zero-point-energy physical state. In this paper, setting up
physical time flow is one of the initial conditions whose physically motivated
appropriate interpretation is central to the purposes of the SRC model. Later,
such physical time reset will also be linked with initial entropy reset. Until
then, we treat time here purely as a coordinate.

\subsection{Evolutionary features at the phase interface}

In Eq. 2.6, the periodic nature of $\delta a\left(  t\right)  $ about the
coasting path , $a_{s}\left(  t\right)  \propto t$,\ implies, for example,
that $\delta a\left(  t_{k-1}=\tau\right)  =\delta a\left(  t_{k}=0\right)  $.
Thus (from the matter sector) the end of the $\left(  k-1\right)  ^{th}$
cosmic kalpa and the beginning of the next $k^{th}$ coincide. This then
suggests that, in general, taking into consideration the entire dynamics.
(including both the contribution due to basic space and that due to matter
fields), cosmic events at $t_{k-1}=\tau\ $and those at $t_{k}=0$ should
coincide, so that both, the end of the $\left(  k-1\right)  ^{th}$ kalpa at
$t_{k-1}=\tau\ $and the beginning of the consequent $k^{th}\ $kalpa at
$t_{k}=0$, as coordinate points in different phases represent the same
temporal instant. If we can demonstrate this consistently, then we can
justifiably move to infer initial physical conditions for the $k^{th}$ kalpa,
predicated on the end-state of the $\left(  k-1\right)  ^{th}$ kalpa. In the
rest of this Section we investigate the soundness of this assertion, and
explore its consequences, if valid.

\subsubsection{Conformal versus non-conformal features}

At $t_{k-1}=\tau$, marking the end of the end of the $\left(  k-1\right)
^{th}$ kalpa, the Hubble parameter (see Eq. 2.8) takes on a simple form with a
boundary value: $H_{\tau}=\frac{1+\cos\psi}{\tau+\tilde{\tau}\sin\psi}$, where
(recall) $\tilde{\tau}=\frac{\tau}{2\pi}$ In [1] we defined the phase constant
$\psi$ in terms of a fundamental interval $t_{f}$ such that $\psi=\frac{2\pi
t_{f}}{\tau}=\frac{t_{f}}{\tilde{\tau}}$, with $t_{f}\lll\tau$. Further, in
[1] we have also estimated $\tau=19.38Gyr$ and taken $t_{f}\sim t_{pl}$
($t_{pl}$ is Planck time) making $\psi$ vanishingly small, $\psi
\simeq5.5\times10^{-61}$. We can therefore (legitimately) set $\cos\psi
\simeq1$ and $\sin\psi\simeq\psi$ so that at $t_{k-1}=\tau$ we effectively
have
\begin{equation}
H_{\tau}=\frac{1+\cos\psi}{\tau+\tilde{\tau}\sin\psi}\bumpeq\frac{2}%
{\tau+t_{f}}\rightarrow\frac{2}{\tau}. \tag{3.1}%
\end{equation}
Further, referring to Eq. 2.8, during the $k^{th}$ kalpa, one finds that at
$t_{k}=0$, $H_{t_{k}=0}\mathbf{=}\frac{1+\cos\psi}{\tilde{\tau}\sin\psi}$, so
that effectively
\begin{equation}
H_{t_{k}=0}=\frac{2}{t_{f}} \tag{3.2}%
\end{equation}
Given that $\left\vert \tau\right\vert \gg\left\vert t_{f}\right\vert $\ it is
clear that the results of the Hubble parameter values in Eqs. 3.1 and 3.2 do
not match. The results do instead imply a discontinuity $\Delta H$, leading to
an energy density discontinuity, in turn.

On the other hand, we also note that the metrics $g_{\mu\nu}^{\left(
k-1\right)  ^{th}}$ and $g_{\mu\nu}^{k^{th}}$ at these points can be
conformally related. \\Thus setting $g_{\mu\nu}\left(  \eta\right)  =\left\{
\exp-\left[  2H_{t_{k}=0}t\right]  \right\}  g_{\mu\nu}^{k^{th}}\left(
t_{k}=0\right)  =\left\{  \exp\left[  -2H_{\tau}t\right]  \right\}  g_{\mu\nu
}^{\left(  k-1\right)  ^{th}}\left(  t_{k-1}=\tau\right)  $, where $\eta$ is a
conformal time, we have that%

\begin{equation}
g_{\mu\nu}^{k^{th}}\mid_{t_{k}=0}=\varphi^{2}g_{\mu\nu}^{\left(  k-1\right)
^{th}}\mid_{t_{k-1}=\tau} \tag{3.3}%
\end{equation}
where now $\varphi=\exp\left(  \frac{1}{t_{f}}-\frac{1}{\tau}\right)  $ is a
conformal factor.

Below we develop arguments that reconcile these seemingly disparate ideas and
use them to develop a physical picture that seeks to evolve the $k^{th}$
cosmic kalpa from its predecessor, the $\left(  k-1\right)  ^{th}$ kalpa.

\subsubsection{Spacetime elasticity and phase transition}

In General Relativity spacetime is dynamic. In cosmology, in particular,
observations indicate space expands (or more specifically stretches), as
originally described by Lamaitre and observed by Hubble. Moreover, in SRC
basic-space is shown to expand monotonically, even in (idealized) absence of
free matter-energy fields, an emergent characteristic that suggests space
should have internal structure [1]. Putting these concepts together leads one
to infer that basic-space displays dynamical characteristics akin to those of
a classical material medium; more specifically characteristics of an elastic
medium. Indeed, this behavior is already manifestly apparent in SRC. According
to DEPP (section 2) basic-space, in its dynamical evolution, is always in
search of dynamic equilibrium as an attractor. In particular, whether under
compressional stresses due to gravitating matter-fields or whether under
stretching stresses due to dark energy's negative pressure-effects,
basic-space's recoil response of $\ddot{a}\propto-a\left(  t\right)  $ in SRC,
is both physically and mathematically indistinguishable from that of an
equilibrium-seeking elastic medium. Thus, here, the matter-energy sector
contribution to cosmic dynamics satisfies a Hooke's law-type relation [1].%

\begin{equation}
\ddot{a}_{m}+\epsilon a_{m}=0, \tag{3.4}%
\end{equation}
for which Eq. 2.5 is clearly a solution, with weighted-elastic-constant
$\epsilon=\left(  \frac{2\pi}{\tau}\right)  ^{2}$.

Now, the concept that space behaves like an elastic medium is not new. Besides
our result in Eq. 3.4, there have been several works alluding to space as an
elastic medium (see e.g. [36][37] [38]). For example, by considering continuum
space as an elastic medium Padmanabhan [36] has derived the Einstein field
equations as consistency conditions, and without the usual need to vary the
metric. Here, Padmanabhan makes an observation that, neither the metric
$g_{\mu\nu}$ nor the energy stress tensor $T_{\mu\nu}$ are fundamental variables.

In our current treatment of SRC, it is seen (Eq. 2.6) that by the time cosmic
dynamics evolves the Universe to the age of $t_{k-1}=\tau$, the latter will
have been monotonically stretched by the dark-energy-driven acceleration since
cosmic age of $t_{k-1}=\frac{1}{2}\left(  \tau-t_{f}\right)  $, for as long as
$\frac{\tau}{2}$ (here estimated to span $\sim9.68Gyrs$). The Universe, which
earlier on would have been quite clumpy, is by then so dilute that it is
virtually spatially flat, $k=0$. In SRC, this acceleration actually ends at
$t_{k-1}=\left(  \tau-t_{f}\right)  $, just before the end of the $\left(
k-1\right)  ^{th}$ kalpa, (recall .$\psi=\frac{2\pi t_{f}}{\tau}$). Here then
both the stretching (dark energy) and space recoil forces just balance so that
$\frac{\ddot{a}}{a}=-\frac{4\pi G}{3}\left(  \rho+3p\right)  =0$ and the
Universe just attains a momentary state of dynamic equilibrium. Therefore, not
only is space now flat, as just pointed out, but $\frac{\ddot{a}}{a}=0$\ means
spacetime also flattens. We discuss these results and more of their
consequences later. Here, we just note as a resulting indication of this
spatial dilution and spatial flatness that such space is by now dominated by
voids of massless fields, only punctuated by scanty structure here and there.
Further, one notes (see Eq. 2.6) that at the final turn of the kalpa, $\left(
\tau-t_{f}\right)  \leq t_{k-1}\leq\tau$, the local dynamics becomes dominated
by the energy in the phase constant $\psi=\frac{2\pi t_{f}}{\tau}$, now the
only operative quantity. Therefore, taking the proposal of space as an elastic
medium further, our interpretation is that at $\tau-t_{f}$ the now maximally
expanding, locally overstretched, basic-space (void) attains its elastic
limit, reaching a critical or yielding point. Here, basic space's
strain-response to stress no longer follows the simple linear Hook's law-type
relation (implied by Eq.3.4). Instead, the expansion takes on a typical
run-away behavior reminiscent of a plastic medium. As a consequence, the
system undergoes a phase transition that breaks the elastic symmetry. One
expects that, as is typical of such critical phenomena, such phase transition
is initially seeded in a small patch of the region. Further, as just observed
above and consistent with the SRC dynamics (Eq. 2.6 see also Eqs. 4.5 and
4.6), this event is associated with huge energy density $\sim\left(  \psi
\tau\right)  ^{-2}$

A general consideration of an isotropic and homogeneous elastic medium
undergoing a phase transition, into a plastic state was previously considered
by [39] as a spontaneous breaking of elastic symmetry. From a different view
point the process just considered can be compared to the pre-inflationary
scenario in quantum cosmology, in which the Universe undergoes a phase
transition from a non-zero ground state vacuum into a false vacuum with higher
potential energy expectation value, just before inflation [40].\textit{ A
general comparison of the similarities and differences in the two scenarios is
taken on later in the discussion of Section 5.}

\subsection{Time-translation symmetry breaking}

The thermodynamic phase transition just highlighted above, initiating at
$t_{k-1}=\left(  \tau-t_{f}\right)  $, is also the driver of the shift we seek
to transit the Universe from one kalpa to another, in this case from the
$\left(  k-1\right)  ^{th}$ kalpa to the next or $k^{th}$ kalpa. Therefore,
going forward, it is crucial to first make physical sense of this phase
transition in the context of the SRC model. Recall that in SRC, cosmic
dynamics implies monotonic space expansion arising, in part, from its
intrinsic (or basic-space-driven) time-linear motion, modulated by
matter-energy-driven harmonic oscillation. The oscillation (Eq. 2.5) spans a
periodic interval $\tau$ that happens to be exactly same as that utilized in
the preceding discussion leading to the phase transition. While such
similarity is not an accident, it is also tempting to infer based on this
similarity alone that end of such an oscillation period at $t_{k-1}=\tau$
unquestionably signals end of the associated physical cosmic time-flow as we
know it. Such an inference would, however, be unjustified and generally
inaccurate. This is because a periodic end $\left(  t=\tau\right)  $ of a
given oscillating system does not necessarily end (or in any way punctuate)
local physical time-flow itself. Instead, end of an oscillation period usually
just marks a specific coordinate interval measure in a referenced continuous
time-flow. On the other hand if, as implied by the SRC model, cosmic evolution
is to be meaningfully represented in terms of kalpas or physical phases of
spacetime, then end of a kalpa must not just mark some coordinate time
interval, but must manifestly end (or at the least punctuate) physical
time-flow. One, then, need explain what a punctuation of physical time would
mean here and, further, one need justify how such punctuation should be
generated. In turn, a process that punctuates (or in some justifiable way
compromises) time flow, will break time translation symmetry.

To set this scenario up we start by observing that a particle of Planck energy
$E=\hbar\nu$ and an associated rest mass $m$, has from Einstein Special
Relativity a total rest/mass-energy $E=mc^{2}$. Such a particle satisfies the
Planck-Einstein relation,
\begin{equation}
mc^{2}=E=\hbar\nu. \tag{3.5}%
\end{equation}
As has previously been pointed out by Penrose [41], the above
(Einstein-Planck) energy relations imply $m\propto\left(  \nu\sim\frac{1}%
{t}\right)  $, where $t$ indicates time-flow, making mass a time-measuring
clock. In turn, this property that masses (and only) constitute clocks implies
that only massive particles, which propagate on timelike geodesics
($dS^{2}\neq0$), are sensitive to physical time-flow. Conversely massless
particles such as photons, propagating on null geodesics ($dS^{2}=0$), are time-flow-insensitive.

Earlier, we found that the phase transition occurring at $t_{k-1}=\tau$ is
preceded by a long time interval, $t=\frac{\tau}{2}$, of dark-energy-driven
cosmic acceleration. During this long-term acceleration the Universe is
monotonically stretched to become spatially very diluted indeed, at the end.
As previously noted this dilution irons out spatial curvature, making space
virtually flat, $k=0$. Further, we note that in the neighborhood of the the
phase transition $\left(  \tau-t_{f}\right)  \leq t_{k-1}\leq\tau$ the
spacetime acceleration $\ddot{a}$ (hence the Ricci curvature scalar) vanishes
$\frac{\ddot{a}}{a}=-\frac{4\pi G}{3}\left(  \rho+3p\right)  =0$. And
consistent with the Friedmann spacetime symmetries, the vanishing Weyl
curvature tensor also means there are no (significant) massless anisotropies
such as gravitational waves. It follows, therefore, that by $\left(
\tau-t_{f}\right)  \leq t_{k-1}\leq\tau$ the space in question is both
conformally and globally flat, and there are no time-measuring clocks.
Consequently, during the events leading to the above-mentioned phase
transition, time-translation symmetry is either broken or at best not
enforceable. It is indeed under these circumstances of a broken
time-translation symmetry that the cosmic phase transition takes place. Below,
we discuss the consequences of this result with regard to initial conditions
for the post-phase-transition cosmic evolution.

\section{Generating cosmic initial conditions from SRC}

Based on the preceding discussion we can now set up initial conditions on
time, space, energy and entropy for the observed Universe, as implied by the
SRC model. We shall also discuss the characteristics of the initial expansion
that follow soon after the phase transition. These initial conditions need
satisfy two constraints. First, consistent with the aims of this work the
conditions should provide a simple, yet consistent basis for evolution of a
universe from a non-singular beginning. Second, such initially
\textit{non-singular} universe should evolve to reproduce features consistent
with the observed Universe. Additionally, starting with such initial
conditions SRC should consistently predict future dynamical evolution of the
observed Universe into the next, or $\left(  k+1\right)  ^{th}$, kalpa. Before
we enter into this itemized discussion of the initial conditions, however, it
is useful to briefly contextualize what is meant (here) by the phrase Big
Bang. For example: what, if anything, is physically quantifiable about it in
this sense? And how does the Big Bang relate to the soutght-for initial
conditions for the Universe?

\subsection{The Big Bang: A first Order Phase Transition}

In a massless void spatial region that is part of the $\left(  k-1\right)
^{th}$ phase of the Universe, a patch forms, during the interval \textit{
}$\left(  \tau-t_{f}\right)  \leq t_{k-1}\leq\tau$. That patch transitions
into the new $k^{th}$ phase Universe at some $t_{k}=t_{pt}$, where $t_{pt}$ is
the time at phase transition in the $k^{th}$ phase (to be justifiably fixed
below)$\ $such that $t_{k-1}=\tau$ and $t_{k}=t_{pt}$ coincide. The process
describes a first order phase transition, that can be quantified in terms of a
Hubble expansion rate as the order parameter. Let us, for convenience, first
define a quantity called the transitional Hubble parameter, $H_{trns}$, which
characterizes the transition at the kalpa interface. Using the general Hubble
parameter expression of Eq. 2.8 and applying the reasoning that led to Eqs.
3.1 and 3.2, it is straightforward to show that $H_{trns}$ is essentially
quantifiable as a step function whose evolution time-rate can be written as,
\begin{equation}
\dot{H}_{trans}\sim-4\left[  \frac{1}{t_{f}^{2}}-\frac{1}{\tau^{2}}\right]
\delta\left(  t_{k-1}-t_{k}\right)  , \tag{4.1}%
\end{equation}
where $\delta\left(  t_{k-1}-t_{k}\right)  $ is a delta function. Reminiscent
of a first order phase transition, as Eq. 4.1 shows, the process manifests a
discontinuity in the order parameter behavior. Moreover, expected, the two
phases are in equilibrium\footnote{This has several consequeses, including,
time-flow "freezing" (which allows reset, see below) and observational
consequences (discussed in Section 6).} during the phase transition \ There
are also, associated changes in energy, volume (space) and entropy (EVE)
reminiscent of a first order phase transition, and which underlie the initial
conditions here, as discussed in the next Section. This phase transition that
takes the Universe from a preceding phase to the emerging one is what, in SRC,
is identified with the Big Bang. Therefore, in this model, the Big Bang begins
the Universe's new cosmic phase and sets the latter's initial conditions at
$t_{k}=t_{pt}=t_{bb}$. Recall $t_{bb}$ is the instant the Big bang occurs
which (here) is coincident with the instant the cosmic phase transition
occurs. In turn, the initial conditions set up the cosmic dynamic evolutionary
process. In this way, in SRC, the Universe does not evolve from a pre-Big Bang
initial singularity, but instead evolves from a previous kalpa (or phase)
through a non-singular process.

\subsection{Time reset and its justification}

We have found that during its final stage, especially during the final
interval \textit{ }$\left(  \tau-t_{f}\right)  \leq t_{k-1}\leq\tau$ the
$\left(  k-1\right)  ^{th}$ kalpa is identified with a spacetime that is
globally flat. In particular, such a spacetime constitutes an environment made
up of massless conformal fields and hence devoid of time-measuring clocks, as
earlier discussed. In this environment time-translation symmetry is
essentially broken. Now, as was shown (Eq. 3.3) the spacetimes just before and
just after the phase transition are conformally similar. And since conformal
transformations render causal structure invariant, this implies the causal
structure of the spacetime immediately after the Big Bang is similar to that
before the Big Bang. Therefore at the immediate end of the Big Bang, spacetime
still constitutes an environment made of massless conformal fields and hence
devoid of time-measuring clocks. In SRC, according to DEPP matter starts to
form soon after this. It follows, then, that initiation of the $k^{th}$ kalpa
from the previous $\left(  k-1\right)  ^{th}$ kalpa also legitimately resets
(physical) time flow $t$ to commence at $t_{k}=0$. This is our temporal
initial condition identifying commencement of the new cosmic phase. An
immediate consequence of this result is that one can now tag the dynamics of
our observable Universe as commencing at a time $t=0$ without running into
pathological complications of an initial singularity. There is \textit{no
initial singularity}, as we show below. Another consequence resulting from
this breaking of time-translation symmetry is that the system's energy is not
conserved. Therefore initial energy conditions need resetting too, as
discussed below.

\subsection{Initial cosmic size and expansion rate}

The reset of initial physical time $t_{k}=t=0$ to kick-start the $k^{th}$
kalpa, established above, allows description of other initial cosmic
parameters. In particular, applying Eq. 2.6 one notes that the new Universe's
phase initiates from a non-vanishing physical spacetime patch, with a scale
size%
\begin{equation}
a_{t_{k}=0}=\upsilon^{\ast}\left[  \tilde{\tau}\sin\psi\right]  \approx
\upsilon^{\ast}t_{f} \tag{4.2}%
\end{equation}
where $\upsilon^{\ast}=\frac{1}{\left[  t_{0}+\tilde{\tau}\sin\left(
\frac{t_{0}}{\tilde{\tau}}+\psi\right)  \right]  }$ is the scale weighting [1]
that sets to unity the current $\left(  t=t_{0}\right)  $ scale factor,
$a_{0}=1$. The last expression $\upsilon^{\ast}t_{f}$ arises from noting that
by definition $\psi=$ $\frac{t_{f}}{\tilde{\tau}}<<1$ and thus from setting
$\sin\psi\simeq\psi$. It is interesting to trace and relate this initial
spacetime patch, as a seed from the previous kalpa. At the phase transition
which takes place at $t_{k-1}=\tau$ in the preceding $\left(  k-1\right)
^{th}$\textit{ }kalpa, the parent scale factor $a_{\tau}$\ (that gives birth
to the $k^{th}$ kalpa) can be written as
\begin{equation}
a_{\tau}=\upsilon^{\ast}\left[  \tau+\tilde{\tau}\sin\psi\right]
\simeq\upsilon^{\ast}\left[  \tau+t_{f}\right]  . \tag{4.3}%
\end{equation}
Indeed, a quick look at the Eq. 4.2 and Eq. 4.3 \ shows that the patch
$\tilde{\tau}\sin\psi$ that becomes the new Universe is manifestly
identifiable inside its parent phase, much like a child inside its mother's
womb, just before birth. Our interpretation is that all information of the
future-to-be $k^{th}$ kalpa is clearly already encoded in the parent $\left(
k-1\right)  ^{th}$ kalpa that produces it, just before the phase
transition\textit{\footnote{Semantically one take the analogy that the new
phase Universe is genetically related to the previous one.}}.

The initial expansion rate of the Universe's new cosmic phase is also finite,
albeit explosive. It is already defined by Eq. 3.2 as $H\left(  0\right)
\simeq\frac{2}{\tilde{\tau}\psi}=\frac{2}{t_{f}}$. Therefore to its
future-to-be inhabitants (studying their past) the new Universe presents as
its initial observable size a Hubble horizon $R_{\left(  t_{f}\right)  }%
=\frac{c}{H_{\left(  t_{f}\right)  }}$ of size
\begin{equation}
R_{\left(  t_{f}\right)  }=\frac{1}{2}c\psi\tilde{\tau}=\frac{1}{2}ct_{f}.
\tag{4.4}%
\end{equation}
This initial initial seed is, in turn, an umbilically extended part of its now
slower expanding larger parent Universe which has a tiny Hubble parameter-size
$H\left(  \tau\right)  =\frac{1+\cos\psi}{\tau+\tilde{\tau}\sin\psi}%
\bumpeq\frac{2}{\tau+t_{f}}\rightarrow\frac{2}{\tau}$ and huge Hubble horizon
size of $R_{\left(  \tau\right)  }\sim\frac{c}{H_{\left(  \tau\right)  }%
}=c\tau$.

Building on experience of thermodynamic phase transitions that often start
with one or more seeds, it is reasonable to expect that the above phase
transition process will initially involve at least one patch, (though it could
even involve several instantly-nucleated patches. It is also possible, though
probably less so (see entropy conditions below) that the transition could
instantaneously involve the whole Hubble sphere $R_{\left(  \tau\right)  }$
(possibly through less usual communication processes such as spacetime
entanglement). In either case, from the perspective of observers, like us, who
end up inside the future of one such initial patch, it does not matter how
many patches are initiated, or even how large the space involved by the phase
transition is initially. This is because the observers' past will always be
restricted to a Universe that (to them) appears to commence its expansion from
an initial Hubble horizon of size $R_{\left(  t_{f}\right)  }\sim c\psi
\tilde{\tau}=ct_{f}$. In general, however, the nucleation process of a small
single patch or probably few such patches is likely much more probable (see
discussions below on constraints from: (i) initial entropy, (ii) initial
imperfect space dilution and (ii) current observations).

Summing up, it is this tiny patch of size $R_{\left(  t_{f}\right)  }=\frac
{1}{2}ct_{f}\sim ct_{pl},$ bounded by a de Sitter-like initially horizon,
$H\left(  0\right)  \simeq\frac{2}{\tilde{\tau}\psi}=\frac{2}{t_{f}}\sim
\frac{1}{t_{pl}}$ that initiates the new ($k^{th}$ phase) Universe at time
$t=t_{k}=0$.

\subsection{\textbf{Initial energy density \ }}

During the accelerating latter half of the $\left(  k-1\right)  ^{th}$ phase,
the Hubble parameter (Eq. 2.8) actually first increases\textit{\footnote{This
is one dynamical difference between SRC and $\Lambda$CDM model [1].} }to a
local maximum $\dot{H}\left(  t_{k-1}\right)  =0$. It then to tapers off, even
though the system continues accelerating to attain its maximum expansion rate
$\ddot{a}\left(  t_{k-1}\right)  =0$ at $t_{k-1}=\left(  \tau-t_{f}\right)  $,
just before the phase transition commences. At this very end of the $\left(
k-1\right)  ^{th}$ phase, as per Eq. 3.1, the system attains a Hubble
parameter value $H_{\tau}=\frac{2}{\tau}$, with associated energy density
$H_{\left(  \tau-t_{f}\right)  }^{2}=\frac{2}{\tau}$given by%

\begin{equation}
\varepsilon_{\tau}=\frac{3}{8\pi G}H_{\left(  \tau-t_{f}\right)  }^{2}%
=\frac{3}{2\pi G\tau^{2}}. \tag{4.5}%
\end{equation}
Following the phase transition, the new $k^{th}$ phase Universe initiates at
$t_{k}=0$ with a Hubble parameter of $H_{t_{k}=0}\simeq\frac{2}{\tilde{\tau
}\psi}=\frac{2}{t_{f}}$, which is associated with energy density
$\varepsilon_{0}$ given as%

\begin{equation}
\varepsilon_{0}=\frac{3}{8\pi G}\left(  \frac{2}{\psi\tilde{\tau}}\right)
^{2}=\frac{3}{2\pi Gt_{f}^{2}}. \tag{4.6}%
\end{equation}
In this way the Big Bang in SRC can be considered as a first order phase
transition, with an associated finite energy change of $\Delta\varepsilon
=\varepsilon_{0}-\varepsilon_{\tau}$ implied between Eq. 4.6 and Eq. 4.5. As
we previously pointed out, the observation that the phase transition process
occurs in an environment in which time-translation symmetry is broken implies
that energy conservation is not guaranteed. This justifies, in part, the huge
energy density $\Delta\varepsilon$ associated with the phase transition. We
shall take the expression of Eq. 4.6 to represent the working initial energy
density of the Universe's new phase.

\subsection{Entropy Initial Conditions}

We now touch on the issue of initial cosmic entropy as a boundary condition,
in the context of SRC model. We start with a brief overview of the current
state of the topic and thereafter inquire on any contributions implied by the SRC\ model.

\subsubsection{Background: The Initial Entropy Problem (IEP)}

The cosmic microwave background CMB radiation produced from the early Universe
is known to be virtually thermal, and hence with a maximal entropy at its
production. But the Universe has evolved since and continues to do so,
suggesting increasing entropy. This apparent contradiction is part of the
Initial Entropy Problem (IEP) [19][42][43][44][45], which part is now
addressed by adding to the thermodynamic (radiation) entropy that associated
with gravitational degrees of freedom, as first suggested by Penrose [46].
Then, there is the other part of the IEP. It arises from the need to explain
how it is that the cosmic initial entropy was low in the first place, given
that there are many more ways to choose a universe under equilibrium
conditions of high entropy [47]. An idea, which has a significant following in
cosmology (see for example [48][49]), is that equilibrium conditions
(including highest entropy) do form boundary conditions for an evolving system
and so the Universe should have initially evolved from such conditions. Thus,
initial cosmic conditions here would involve some dynamical mechanism by which
a low entropy fluctuation evolves from a background high entropy state [50].
The interpretation then [45] is that initial conditions should not be assumed
a priori, but should rather be generically set up by existing physics. Such
thinking is consistent with the Boltzmann Anthropic Hypothesis [51][52], in
which Boltzman first proposed that the low entropy of the Universe should be a
random fluctuation from a maximal entropy state. The fluctuation theorem [50]
implies that small fluctuations from highest equilibrium entropy are
exponentially more likely in this scenario than large fluctuations. On the
other hand, proponents of inflation for cosmic initial conditions point out
that a low entropy initial condition, as in the Past Hypothesis, is favored
over a high entropy one by virtue of the Second Law of Thermodynamics [45]
[53]. Indeed, the absence of such a (Boltzmann) background high entropy pool
[51][52], along with the lack of a dynamical mechanism by which a low entropy
fluctuation [50] could evolve from such a high entropy pool (on the one hand),
and the lack of a basic principle on which to ground the Past Hypothesis (on
the other hand), are both at the base of the other (or remaining) part of the
initial entropy problem (IEP).

More generally, at issue appears to be a subtle but deeper and potentially
intricate argument of boundary conditions. Proponents of the Boltzman
Hypothesis, who argue that low initial entropy should be explained from a low
entropy fluctuation originating from a high entropy equilibrium background,
point to the logic for upholding the concept of time-symmetry, implied in
physical laws [49]. Proponents of inflation for initial conditions, on the
other hand [42] [45] [53], point out that the Past Hypothesis and the Second
Law of Thermodynamics suggest unidirectional, time-asymmetric cosmic
evolution. What makes the argument even much more challenging is the fact that
as physicists we are all trained implicitly to subscribe to both the logic of
time-symmetry as implied in physical laws, and to the logic of time asymmetry
and the arrow of time as implied by Second Law of Thermodynamics, for example.
Framed this way, this dichotomy in time symmetry makes the IEP likely one of
the most outstanding unresolved issues in physics, probably side-by-side with
the Cosmological Constant Problem.

\subsubsection{Entropy at the end of a kalpa in SRC}

In the previous Section we have considered the evolution of the Universe
during a given kalpa, the $\left(  k-1\right)  ^{th}$ kalpa, spanning $0\leq
t_{k-1}\leq\tau$. We argued that at $t_{k-1}=\tau-t_{f}$ the
dark-energy-driven cosmic acceleration ends and a phase transition sets in to
initiate the next (or $k^{th}$) kalpa (which for our Universe we identify with
the current phase). Based on statistical considerations, we now inquire on the
likelihood of the SRC phase transition process leading to a new phase, one
consistent with our observed Universe. According to SRC, at the end of the
$\left(  k-1\right)  ^{th}$ phase when $\ddot{a}\left(  t_{k-1}\right)  =0$
(just before the assumed phase transition) the Universe attains both dynamic
and thermodynamic equilibrium. Here, the Universe therefore has maximum
equilibrium entropy. Our plan will be to show that these equilibrium
conditions (including highest entropy) do form boundary conditions from which
the Universe's new cosmic phase initially evolves. At issue then will be to
find (as suggested [50]) some dynamical mechanism by which a low entropy
fluctuation evolves from such background high entropy state, and to directly
identify such a fluctuation with the Universe's new cosmic phase.

Following Gibbons and Hawking, [54][48], one can write down the maximum
entropy enclosed by this horizon at $\ddot{a}\left(  t_{k-1}\right)  =0$ as%

\begin{equation}
S_{\left(  \tau\right)  }=k_{B}\pi\frac{R_{\left(  \tau-t_{f}\right)  }^{2}%
}{l_{pl}^{2}}, \tag{4.7}%
\end{equation}
where $k_{B}$ and $l_{pl}$ are the Boltzman constant and Planck length,
respectively and where.%
\begin{equation}
R_{\left(  \tau-t_{f}\right)  }=\frac{c}{H_{\left(  \tau-t_{f}\right)  }%
}=\frac{1}{2}\left(  \tau-t_{f}\right)  c\simeq\frac{1}{2}\tau c. \tag{4.8}%
\end{equation}
The maximum entropy result in Eq. 4.7 is idealized since, as we have
previously noted at the end of its $\left(  k-1\right)  ^{th}$ kalpa, the
Universe while very dilute in its final state, still contains some lumpy
structures in the form of unradiated black holes and possibly galaxies. Still,
one can estimate the combined net cosmic entropy in this final state. In fact,
Gibbons and Hawking [54] gave a prescription on calculation of combined
entropy, here referred to as $S_{c}$, of such a horizon-bounded configuration
which contains extra matter. They found that introduction of localized
structure or even radiation inside such horizon depresses the horizon entropy,
as it reduces the horizon size. In this case, the result is%

\begin{equation}
S_{c}=\left[  S_{\left(  \tau\right)  }-\sqrt{S_{b}S_{\left(  \tau\right)  }%
}+S_{b}\right]  , \tag{4.9}%
\end{equation}
where $S_{b}$ is the entropy of a black hole with ADM mass equivalent to any
such localized matter. Note that consistently here $S_{c}<S_{\left(
\tau\right)  }$, so that $S_{\left(  \tau\right)  }$ is maximal. We will take
Eq. 4.9 to represent the net entropy at the end of the cosmic phase. However,
for purposes of the discussion here we shall, with no loss of generality,
still apply the maximal entropy $S_{\left(  \tau\right)  }$ of Eq. 4.7 as the
background equilibrium entropy.

\subsubsection{Seeking a resolution to the IEP from SRC}

Recall that at (and immediately after) the phase transition, the resulting
patch that initiates the new kalpa at $t_{k}=0$ is dominated by the energy in
the phase constant, Eq. 4.6. In particular (from Eq. 2.6 or Eq. 4.2) the
Hubble parameter here is flat, $\dot{H}_{\left(  t_{k}=0\right)  }=0$, and the
initial boundary of the new patch is momentarily de Sitter-like. We shall
later comment further on the consequential dynamical implications of this
result later. For now, we note [54][48] that one can therefore write the
horizon entropy of this nucleated post-transition patch as
\begin{equation}
S_{\left(  t_{f}\right)  }=k_{B}\pi\frac{R_{\left(  t_{f}\right)  }^{2}%
}{l_{pl}^{2}}, \tag{4.10}%
\end{equation}
where $R_{\left(  t_{f}\right)  }^{2}$ is given by Eq. 4.4. It should be
re-emphasized here that the patch at which the phase transition initiates the
Universe's new $k^{th}$ phase is umbilically connected to (or is a piece of)
the outgoing $\left(  k-1\right)  ^{th}$ phase. Therefore considering that
during the transition process the entropy $S_{\left(  t_{f}\right)  }$ of the
patch is linked to the original background $R_{\left(  \tau\right)  }$, with
its equilibrium entropy $S_{\left(  \tau\right)  }$,\ and applying the
Gibbons-Hawking approach as in Eq. 4.9, the resulting combined entropy
$S_{\left(  rc\right)  }$ is%

\begin{equation}
S_{\left(  rc\right)  }=\left[  S_{\left(  \tau\right)  }-\sqrt{S_{\left(
t_{f}\right)  }S_{\left(  \tau\right)  }}\right]  , \tag{4.11}%
\end{equation}
where (compared with Eq. 4.11) we can neglect the last term in $S_{\left(
t_{f}\right)  }$ on the basis that $\frac{R_{\left(  t_{f}\right)  }%
}{R_{\left(  \tau\right)  }}\lll1$. It follows from Eq. 4.11 that the patch
entropy in Eq. 4.9, $S_{\left(  t_{f}\right)  }\ll S_{\left(  rc\right)  }$,
emerges as a fluctuation of the background entropy, $S_{\left(  \tau\right)
}$ in Eq. 4.7.

Following [48] one can set the probability for such a fluctuation $P_{\left(
t_{f}\right)  }$, and by implication the probability for the phase transition,
as
\begin{equation}
P_{\left(  t_{f}\right)  }=\exp\left(  -\sqrt{S_{\left(  t_{f}\right)
}S_{\left(  \tau\right)  }}\right)  , \tag{4.12}%
\end{equation}
where $S_{\left(  \tau\right)  }$ and $S_{\left(  t_{f}\right)  }$ are given
by Eqs. 4.7 and 4.9, respectively. Thus, for a fixed equilibrium background
entropy (such as $S_{\left(  \tau\right)  }$ is here) clearly the smaller
$S_{\left(  t_{f}\right)  }$ the larger its probability $P_{\left(
t_{f}\right)  }$ is. Still, at the same time, the result in Eq. 4.9 supports
the idea advanced by Albrecht [48] that the small initial cosmic entropy (here
$S_{\left(  t_{f}\right)  }$) should, nevertheless, be de Sitter-maximal.

These results can now be viewed, first in the context of addressing the IEP
and second in terms of ability for SRC model to construct initial conditions
that can reproduce the observed Universe. The result in Eq. 4.12 is consistent
with the demands of the Fluctuation Theorem [50] that small entropy
fluctuations about the background entropy are exponentially more likely than
large ones. In particular, given that $S_{\left(  t_{f}\right)  }\ll\left[
S_{\left(  rc\right)  }\lessapprox S_{\left(  \tau\right)  }\right]  $, the
resulting large probability $P_{\left(  t_{f}\right)  }$ implies that the
phase transition process, as depicted by SRC is indeed a favored process for
generating appropriate initial entropy conditions. More generally, the
observation that the patch that initiates the Universe's new $k^{th}$ phase is
a piece of the preceding $\left(  k-1\right)  ^{th}$ phase, and that its
initial entropy $S_{\left(  t_{f}\right)  }$ (in Eq. 4.9) is therefore a
fluctuation of the latter's background entropy $S_{\left(  \tau\right)  }$, is
consistent with some previously well-established views. First, the findings
are consistent with with Penrose's view [47] [45] that such initial conditions
should not be fixed a priori, as implied by the Past Hypothesis [42] [45]
[53], but that such conditions should instead generically emerge from
pre-existing physics [56]. Secondly, the results are consistent with the
Boltzmann Anthropic Hypothesis [51][52] that the initial low entropy of the
Universe should be a random fluctuation from a maximal entropy state. As
previously pointed out in [45] the Boltzmann's Anthropic Hypothesis, is itself
an example of the Poincare Recurrence Theorem [55]. This latter theorem
implies that, given enough time, a system will eventually return arbitrarily
close to its initial state in phase space.

Apparently, the issue of recurrence as a general evolutionary feature of the
Universe is already implied in SRC, through the manifested periodicity of
cosmic expansion in general (Eq. 2.6) and particularly through the periodic
breaking of time-translation symmetry that renders kalpa-to-kalpa
time-resetting possible, as discussed in the previous section. Therefore, in
SRC the Poincare Recurrence Time for the Universe corresponds to a kalpa
duration, $\tau$. The results appear to provide the needed dynamical mechanism
[47] by which, as part of initial cosmic conditions, a low entropy fluctuation
(here $S_{\left(  t_{f}\right)  }$) could have evolved from a
background\ state with high equilibrium entropy $S_{\left(  \tau\right)  }$.
It is in this context, as the results imply, that SRC does offer a means to
address the IEP.

On the other hand, one also observes that (locally) within a given kalpa in
SRC the entropy evolution bounded by the evolving Hubble horizon is expectably
a monotonically increasing quantity, beginning as $S_{\left(  t_{f}\right)  }$
at the start of the Universe's new cosmic phase and ending at the phase-end as
$S_{\left(  \tau\right)  }\gg S_{\left(  t_{f}\right)  }$. Such evolution is
consistent with the Second Law, which this model also necessarily upholds,
along with the implied future-directed time arrow. Further, we have
demonstrated that at the beginning of each kalpa, the Universe starts off at
its lowest entropy $S_{\left(  t_{f}\right)  }$. While this reads like a
restatement of the Past Hypothesis, note that the SRC\ results also go beyond
the Past Hypothesis to explain the source of the initial low entropy, as a
fluctuation of a larger equilibrium entropy of the past phase.

In summary, the outgoing discussion suggests SRC can contribute towards a
resolution of the Initial Entropy Problem. It provides appropriate initial
entropy conditions, requiring that the initial low entropy of the Universe be
a random fluctuation from a maximal entropy state. As a dynamical system the
Universe is seen to obey the Poincare Recurrence Theorem [55], from kalpa to
kalpa, with a well-defined Poincare Recurrence Time $\tau$, as described by
its evolutionary equations for $a\left(  t\right)  $, $\dot{a}\left(
t\right)  $ and $\ddot{a}\left(  t\right)  $ (see e.g. Eq. 2.6 and Eq. 3.4)
and by DEPP [1]. This recurrence does not suggest here a process upholding
time-reversal symmetry since, indeed, cosmic expansion in SRC is monotonic. On
the other hand, locally within a given phase $\left(  0<t_{k}<\tau\right)  $
cosmic evolution upholds the Second Law of Thermodynamics. Thus, within a
given kalpa, the concept of a future-directed time-arrow is in force. These
findings may suggest cosmic dynamics as an interplay of two time scales, a
universal kalpa-to-kalpa recurrent one, and a local and future-directed
asymmetric one. Finally, once again, consistent with the demands of the
Fluctuation Theorem [50] and \ given that $S_{\left(  t_{f}\right)  }%
\ll\left[  S_{\left(  rc\right)  }\lessapprox S_{\left(  \tau\right)
}\right]  $, the resulting large probability $P_{\left(  t_{f}\right)  }$ (Eq.
4.12) implies the Big Bang phase transition process as depicted by SRC is
indeed a favored process of cosmic evolution. The findings also imply SRC
creates cosmological scenario in which initial conditions [47] are predicted
by the dynamics [56], and not merely assumed. A detailed discussion on this
topic will appear elsewhere.

\section{Comment on preliminary cosmic evolution after phase transition}

In the preceding discussion we have described pre-evolutionary conditions
leading to the Big Bang in SRC, discussed the Big Bang process as a first
order phase transition and set up initial conditions for the post-Big Bang
evolution. These include initial temporal, spatial, energy and entropy
conditions. Given such initial conditions how then, one can inquire, does
cosmic dynamics proceed immediately after the phase transition? What is the
starting dynamical character of the Universe's new cosmic phase? Does such
dynamics demonstrably follow from the set initial conditions? And does such
initial dynamics predict the observed Universe? Answers to these questions can
contribute to indicators of whether or not the SRC model can be taken seriously.

\subsection{Horizon and flatness issues}

We have previously shown that as a consequence of the extended period of
cosmic acceleration which leads to the end of the $\left(  k-1\right)  ^{th}$
space undergoes increasingly extensive dilution. Thus by the time the Big Bang
phase transition into the $k^{th}$ kalpa occurs any previous spatial curvature
has unwrinkled out, $k=0$ and space is virtually flat. Moreover, based on the
evolution of the scale factor (Eq. 2.5), the cosmic acceleration ends at
$t_{k-1}=\left(  \tau-t_{f}\right)  $ and the zero-mass-condition $\frac
{\ddot{a}}{a}=-\frac{4\pi G}{3}\left(  \rho+3p\right)  =0$ holds so that at
this point spacetime is all flat. This last characteristic suggests spacetime,
before the Big Bang, is in thermodynamic equilibrium. Recall also, as
previously shown in Eq. 3.3, the spacetimes just before and just after the
phase transition are conformally related. And since conformal re-scaling
preserves the causal structure, as light cones remain invariant, both
characteristics of flatness and thermodynamic equilibrium are still in force
just after the phase transition. Thus, according to the SRC model, the new
$k^{th}$-phase Universe has neither a flatness nor a horizon problem at
$t_{k}=0$. Clearly, while this result does not rule out inflation, it makes
moot the reasons usually used to justify it and introduce it by hand as in
some models. As we point out below, though, SRC has its surprises.

\subsection{Inflation or no inflation?}

At the exit of the phase transition, referred to as the Big Bang here,
$t_{k}=0$, the patch that becomes the new (or $k^{th}$) phase Universe is
finite as indicated by Eq. 4.4, and with a well-defined Hubble horizon. We
note again (based on Eq. 2.5) that during the initial period that follows
$t_{k}=\succeq0$ the dynamics is dominated by the effects of the phase
constant $\psi$. This phase constant manifests characteristics reminiscent of
the cosmological constant $\Lambda$, in $\Lambda CDM$. In particular one can
relate the two such that $\Lambda\approx\left(  \frac{2\pi}{\tau\psi}\right)
^{2}$. Writing down the scale factor as $a\left(  t\right)  =a_{0}\exp\left(
\int Hdt\right)  $, it is seen (Eq. 3.4) that Hubble parameter is flat at
$t_{k}=0$, (and virtually so in the immediate neighborhood). Therefore here
one has that $a\left(  t\right)  =\exp\left[  H_{\left(  t_{f}\right)
}t\right]  $ or
\begin{equation}
a\left(  t\right)  =\exp\left[  \left(  \frac{4\pi}{\psi\tau}\right)
t\right]  =\exp\left[  \left(  \frac{2}{t_{f}}\right)  t\right]  . \tag{5.1}%
\end{equation}
Thus for a brief period $\succsim t_{f}$ this huge and virtually flat Hubble
parameter, with a correspondingly huge amount of energy density $\varepsilon
_{\max}$ (Eq. 4.6) renders the dynamics de Sitter-like. It follows then, that
according to the SRC model, the initial period of the Universe's new cosmic
phase is naturally inflationary.

We emphasize that this inflationary feature is natural in the model and is not
introduced to solve any existing horizon or flatness problems. We saw earlier
on that in the model, the new phase commences with neither a flatness nor a
horizon problem. Still such inflationary period can have a role to play in,
for example amplifying any quantum fluctuations generated during the phase
transition into scalar or density perturbations that seed future structure
formation. A quantified discussion of these features will take place elsewhere.

\subsubsection{Reheat}

Inflation is non-equilibrium dynamics and in SRC the need to regulate cosmic
dynamics towards equilibrium, demanded by DEPP (Section 2), implies a
deceleration to exit inflation. The deceleration sets in soon as $\psi$ no
longer dominates the dynamics, when $t>t_{f}$. To effect the needed
deceleration and slow down toward equilibrium, the Universe increases its
inertia through creation of matter-energy. In this way, cosmic inflation which
in SRC emerges naturally also finds a graceful exit, through matter creation
according to DEPP. Once again, inflation and its graceful exit here are both
achieved with no need to invoke a singularity-sourced scalar field and a
fine-tuned potential for the purpose. This is reheat, and thereafter the
Universe develops as observed.\textbf{ }As with the inflationary features
above, a quantified discussion of the reheat period and matter creation will
follow elsewhere.

\section{Potential Tests of SRC: Mature Structures at High Redshits}

Before closing this paper it is reasonable to ask whether there could be
features or phenomena the model could uniquely predict and/or explain. Recall
that the original patch that becomes the new-phase Universe at phase
transition, originated as part of the background in the previous phase. Once
this patch starts to inflate, it will appear to a distant future observer
inside it to have causally disconnected from any previous phase background.
This is because the de Sitter-like inflationary Hubble horizon (with a very
high Hubble parameter) is at the time highly constricted for the inside
observer in its future. Indeed the notion of a past cosmic phase for such an
observer, like us, is not straightforward. In a global sense, though, this
patch that drives into the new phase is always part of the extended cosmic
spacetime. This is one characteristic feature of the Big Bang in SRC model
that is absent in other models, including eternally inflating models.
Therefore, in SRC the new patch appears like a new separate Universe, only for
the future observer inside it, restricted by its horizon. However, as
previously pointed out, the inflationary phase lasts a short period, and soon
the inevitable inertia-induced deceleration sourced by matter-energy creation
(see DEPP and Eq. 2.5) sets in. A consequence of this inertia-induced
deceleration phase is\ the eventual disappearance of the de Sitter-like
horizon that previously causally separated the new-phase inflationary patch
from the background during, and right after, the (Big Bang) phase transition.
This leaves behind a regular (non-de Sitter) Hubble horizon. A related
consequence of the growing deceleration is the corresponding widening of the
Hubble horizon. In particular, as more and more matter is created and this
space becomes more and more gravitationally dominated, eventually the
new-phase sector is expanding at about the same rate (or even slower) than its
original past-phase immediate neighborhood. This widens the Hubble horizon
into the old-phase background, with the result that old structures from the
previous phase (such as galaxies and black holes) can cross the growing
Hubble\ horizon, into the observable view of the Universe's new cosmic phase
future observer. This is a prediction of the SRC, completely consistent with
the preceding analysis.

The first of the structures to enter the horizon would be those which
previously were within the neighborhood of the patch site that formed the new
phase now containing the observer. Future observers could distinguish such
"alien" structures (coming from the past phase) from those other large-scale
structures actually formed in the contemporary new phase. In particular, those
past phase structures that crossed the horizon early during the Dark Ages
(before the expected commencement of the large-scale structure formation
period of the Universe's new cosmic phase commenced) would appear relatively
out-of-place, as galaxies or black holes too mature for that cosmic era [58].
Figure 2 is a diagrammatic sketch of cosmic evolution in comoving coordinates
(based on $\Lambda CDM$\textit{\footnote{For ease of communicating our point,
we have chosen to use the familiar $\Lambda CDM$ cosmic horizon evolution
diagram with no loss of generality. The SRC (not shown here) appears slightly
different mostly around the matter-dark energy equality line about 4.07 Gyrs
ago.})}.We have modified \ the original diagram (\textit{Courtesy of \ Physics
Stack Exchange}) to add features identifiable with the past cosmic phase. The
solid baseline represents the extent of the Hubble horizon of the past phase
at the instant of the phase transition. The phase transition (Big Bang) takes
place at the cosmic past-phase/future-phase interface, marked here by
intersection of the solid baseline with the observer's (solid vertical)
worldline. The dotted vertical lines parallel to the solid line show comoving
worldlines of "alien" structures (galaxies/black holes) from the past phase.
Each of the arrows indicates the instant of entry of the associated structure
into the Hubble sphere (oval region) through the Hubble horizon of the
contemporary (current) phase of the Universe. The structures are observed at
their crossing of the observer's past light cone.%

\begin{figure}[ptb]%
\centering
\includegraphics[
height=2.437in,
width=5.1984in
]%
{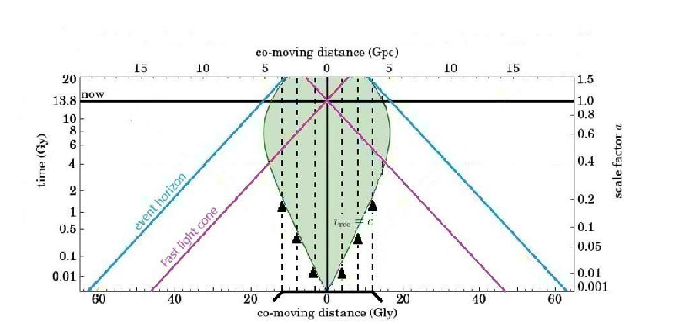}%
\caption{\textit{A sketch of cosmic evolution in comoving coordinates in based
on }$\Lambda CDM$\textit{ (Courtesy of \ Physics Stack Exchange) modified to
describe the process.} \textit{The outline of the oval region shows the
evolution of the Hubble horizon.} \textit{The vertical solid line is the
observer's worldline}. \textit{The solid baseline shows the Hubble sphere size
of past cosmic phase at phase transition. The dotted lines show comoving
structures from the previous cosmic phase entering (at the arrows) the Hubble
horizon of the new cosmic phase.}}%
\end{figure}

The scenario just depicted in above paragraph is reminiscent of recent
observations by the James Webb Space Telescope (JWST). Lately the JWST has
consistently brought in pictures, interpreted to be those of unusually mature
structures at high redshifts, such as mature galaxies [58][59] and mature
black holes [60]. Based on the outgoing analysis, our projection, is that some
of these mature structures observed by JWST at high redshifts should be
structures that entered the cosmic horizon from the previous kalpa or phase of
the Universe. Therefore, identification of such unusually mature structures in
the early universe, as structures from the past phase, could serve to validate
the SRC model.

\section{Discussion}

In regard to cosmic initial conditions, standard cosmology posits that the
Universe originates from a space-like initial singularity, and thereafter
inflates [15] and reheats to create matter and eventually evolve into the
observed state. There are several challenges with the initial part of this
picture. First there is no known prescription in physics on how to evolve a
universe (or any space for that matter) from such a physical singularity into
a geodetically complete spacetime. Secondly, it is even harder to find a
physical principle that evolves an inflationary phase followed by a reheat
phase from such a singularity origin. Much more generally, at issue is whether
conditions of the early Universe can be predicted from some physical
principle. In the past several leading researchers [47][49] have suggested
such cosmic initial conditions should not be fixed, a priori, but instead
should generically emerge from existing physics.\textit{ }In this paper we
have presented an alternative approach to discussing cosmic dynamic and
thermodynamic initial conditions, the evolution of which can lead to the
observed Universe justifiably. The approach is based on a framework provided
by the SRC model [1]. In the SRC model, cosmic dynamics is self-regulating
allowing the Universe to evolve through self-regenerating cosmic phases or
kalpas. Here, the end of a given kalpa and the beginning of the next are
linked by an interface that sets initial conditions for this latter new phase.
The main aim of the paper was to construct this cosmic phase interface and
thereafter discus its role in initiating the new cosmic kalpa. To test its
effectiveness we have used the model to discuss some standing observational issues.

In this closing discussion we recap some features of this treatment. Applying
a course-grained, classical, approach we have with justification [1] [36],
considered space to behave as an elastic material medium with an associated
weigted elastic constant $\kappa\sim\frac{1}{\tau^{2}}$, where $\tau$ is the
duration of a cosmic phase. Taking the analogy further, we noted that at
$t=\tau-t_{f}$, when cosmic acceleration ends just as the kalpa is ending,
this space elasticity reaches a yielding point to undergo a phase transition
thereafter. The phase transition is itself initiated through localized
spacetime sites with nucleated defects (extreme void), so that which the
spacetime "elastic strain" no longer follows a simple linear Hook's law-type
relation. Instead the defects generate de Sitter-like patches so that
space\ here becomes inflationary. From the point of view of an observer inside
the spacetime defect patch's future a new spacetime is created. The new
spacetime's expansion horizon is made initially small\footnote{This is
independent of the actual physical size of this patch.} by the associated
extreme expansion (Hubble) rate, $H_{\left(  t_{f}\right)  }\sim\frac{1}%
{t_{f}}$, of the the patch. This transition of space between two horizons,
from the old phase horizon $R_{\left(  \tau\right)  }=c\tau$ to a new phase
with horizon $R_{\left(  t_{f}\right)  }=ct_{f}$ constitutes, the Big Bang in
SRC. The process which mimics a first order phase transition, also resets time
and seeds all the initial conditions for the new Universe phase, including
energy, size and entropy. The model predicts inflation to follow the Big Bang
(phase transition), naturally, and further, shows how inflation ends
naturally, through matter creation predicted by SRC to facilitate the
equilibrium-seeking decelerations. Hereafter the Universe becomes radiation
dominated, consistent with CMB observations. As a consequence there is no need
to invoke some arbitrary inflaton scalar field, in SRC. The initial conditions
of the new cosmic dynamical phase evolve from the previous phase.

The paper also discussed the issue of initial cosmic entropy as part of
initial conditions. We have shown how the evolution of the Universe in SRC is
based on self-regenerating cosmic phases or kalpas. It satisfies the Poincare
Recurrence Theorem with a Poincare Recurrence Time given by the kalpa interval
$\tau$. We showed that at the inter-kalpa phase transition the Boltzman's
Anthropic Hypothesis, which is an example of Poincare Recurrence Theorem,
holds and the Universe's new cosmic phase's initial low entropy evolves as a
fluctuation from the higher entropy of the outgoing phase. Thereafter, within
the new phase, the entropy evolves consistent with the Second Law of
Thermodynamics. Consequently, while the Boltzman Hypothesis, under the
umbrella of Poincare Recurrence operates globally between kalpas, the Past
Hypothesis and Second Law operate locally within a kalpa. In our view this
reconciles the Boltzman's Anthropic Hypothesis with the Second Law of
Thermodynamics, in the model. We are not aware of a previous such attempt at
this sought-for reconciliation.

The paper closes suggesting one potential signature for a kalpa-based cosmic
self regeneration. Astrophysical structures such as galaxies or black holes,
from previous phases can eventually cross the Hubble horizon of Universe's new
cosmic phase, to be observable. We have therefore conjectured that the
JWST-observed mature structures such as old spiral galaxies and supermassive
black holes at high red-shifts are likely come from the previous phase Universe.

Summarizing the outcomes, we have: (i) set up a theoretical framework based on
[1] and in which space behaves as an elastic medium; (ii) created a
phase-based Universe with 2 horizons, and (iii) argued for a phase transition
between the horizons which we identify as the Big Bang. (iv) As a first order
phase transition the Big Bang admits cosmic initial conditions in terms of
energy, size, entropy (EVE). (v) These initial conditions imply a non-singular
cosmic origin. (v) We have confirmed the beginning of the cosmic phase (kalpa)
to be inflationary. Thus inflation is naturally predicted in SRC. (vi) We have
shown that the end of inflation and matter creation are inevitable
consequences in SRC predicted by DEPP. (vii) We have addressed the Initial
Entropy Problem (IEP), showing how (consistent with Boltzman) initial cosmic
entropy originates as a small fluctuation from large (equilibrium) entropy at
end of the preceding kalpa, and how thereafter within each kalpa cosmic
entropy is consistent with the Second law. Finally:we have used the SRC model
suggest the origin of some over-mature structures observed by JWST.

Assuming the SRC model correctly interprets the observed Universe, there is
room to discuss the processes in more detail, including those that need a
micro-physical level framework.

\section*{Acknowledgements}
I\textit{ would like to thank Edwin Gatwaza for
diagram edits and Albert Munyeshyaka for style editing. This work was
supported by the Swedish International Development Cooperation Agency (SIDA)
through the International Science Program (ISP) grant N\b{o}\ RWA:01.}

\bibliographystyle{iopart-num}
\providecommand{\newblock}{}

  \end{document}